\documentclass[a4paper,11pt]{article}
\pdfoutput=1 

\usepackage{jheppub} 

\usepackage{dcolumn}
\usepackage{graphicx}
\usepackage{epsfig}
\usepackage{hyperref}
\usepackage{graphicx}
\usepackage{dcolumn}
\usepackage{amssymb,eucal}
\usepackage{longtable}
\usepackage{amsmath,amssymb,bm}
\usepackage{mathrsfs}

\newcommand{\mathd}{\mathrm{d}}

\newcommand{\mathe}{\mathrm{e}}

\newcommand\tr{{\rm tr }}

\def\tr{\mathop{\rm tr}\nolimits}

\title{Thermodynamics of large-$N$ gauge theories on a sphere: weak versus strong coupling}
\author[1]{Fen Zuo,}
\author[2]{Yi-Hong Gao}
\affiliation[1]{School of Physics, Huazhong University of Science and Technology, Wuhan 430074, China}
\affiliation[2]{State Key Laboratory of Theoretical Physics, Institute of Theoretical Physics,
Chinese Academy of Sciences, P.O. Box 2735, Beijing 100190, China}
\emailAdd{zuofen@hust.edu.cn}
\emailAdd{gaoyh@itp.ac.cn}
\abstract{
Recently lattice simulation in pure Yang-Mills theory exposes significant quadratic corrections for both the thermodynamic quantities and the renormalized Polyakov loop in the deconfined phase. These terms are previously found to appear naturally for ${\mathcal N}=4$ Super Yang-Mills~(SYM) on a sphere at strong coupling, through the gauge/gravity duality. Here we extend the investigation to the weak coupling regime, and for general large-$N$ gauge theories. Employing the matrix model description, we find some novel behavior in the deconfined phase, which is not noticed in the literature. Due to the non-uniform eigenvalue distribution of the holonomy around the time circle, the deviation of the Polyakov loop from one starts from $1/T^3$ instead of $1/T^2$. Such a power is fixed by the space dimension and do not change with different theories. This statement is also true when perturbative corrections to the single-particle partition functions are included. The corrections to the Polyakov loop and higher moments of the distribution function combine to give a universal term, $T/4$, in the free energy. These differences between the weak and strong coupling regime could be easily explained if a strong/weak coupling phase transition occurs in the deconfined phase of large-$N$ gauge theories on a compact manifold.

}


\begin{document}
\maketitle
\flushbottom

\section{Introduction}
Recent lattice simulation have provided us some novel results in the deconfined phase of pure SU($N$) Yang-Mill theory. In the high temperature region, the thermodynamics approaches the free limit, with the deviation well described by the perturbative contribution. When temperature goes down to $T_c\lesssim T\lesssim 4 T_c$ with $T_c$ the deconfinement temperature, the deviation increases sharply and is dominated by $1/T^2$ corrections. Such a behavior is quite manifest in the plot of the trace anomaly, which vanishes in the high temperature limit. The lattice data for the trace anomaly in SU(3) gauge theory \cite{Boyd:1996bx} shows a clear linear dependence on $T^2$ in such a temperature region~\cite{Pisarski:2006yk}. This behavior is further confirmed for SU($N$) gauge theory with various $N$, and believed to hold in the large $N$ limit~\cite{Panero:2009tv}. The simulation in \cite{Boyd:1996bx} is also extended to a broad temperature region and compared to the perturbative results~\cite{Borsanyi:2012ve}. The comparison shows clearly that the quadratic corrections can not be generated from the perturbative approach. A fuzzy bag model is proposed based on this, in which the pressure is given as an expansion in powers of $1/T^2$. Similar behavior is found for the Polyakov loop, the order parameter for the deconfinement phase transition. In the high temperature limit, the quark free energy goes to zero and the renormalized Polyakov loop approaches one. As temperature decreases, perturbative contributions give a small negative correction to the quark free energy. Accordingly, the Polyakov loop increases slightly above one, which is indeed seen in lattice data. When temperature goes down to a few times $T_c$, the Polyakov loop decreases quickly below one. The logarithm of the Polyakov loop in this region is fitted well with a single $1/T^2$ term~\cite{Megias:2005ve}, with the coefficient depends mildly on the gauge group~\cite{Mykkanen:2012ri}. At the phase transition, the Polyakov loop arrives at a value around one half, for SU($N$) gauge theory with $N=3$~\cite{Kaczmarek:2002mc,Gupta:2007ax} and $N=4, 5$~\cite{Mykkanen:2012ri}.

The gauge coupling is supposed to be strong close to $T_c$, so it is quite suitable to study such power corrections with the gauge/string duality~\cite{Maldacena:1997re,Gubser:1998bc,Witten:1998qj}. The original correspondence is between type IIB superstring theory in 5 dimensional Anti-de Sitter~(AdS) space and ${\mathcal N}=4$ Super Yang-Mills~(SYM) on the boundary Minkowski space. Since the boundary theory is a conformal theory, no bound states exist and the theory is always unconfining at finite temperature~\cite{Witten:1998zw}. The free energy are found to be $3/4$ of that in the zero coupling limit~\cite{Gubser:1996de}. Further calculation shows that such a difference is diminished by the corrections both at strong \cite{Gubser:1998nz} and weak coupling~\cite{Fotopoulos:1998es}. However, it is argued that it is impossible to smoothly extrapolate from the weak coupling to the strong coupling regime, and a phase transition in the coupling must occur~\cite{Li:1998kd}. Later this kind of strong/weak coupling transition is extended to general maximal supersymmetric gauge theories~\cite{Gao:1998ww}, and the transition point is argued to be related to the correspondence point~\cite{Horowitz:1996nw}. An example of strong/weak phase transition is previously found in 2-dimensional lattice gauge theory at large $N$~\cite{Gross:1980he,Wadia:2012fr}. The development of the localization method~\cite{Pestun:2007rz} in recent years makes it possible to examine various theories in the whole coupling region. Interestingly, a series of strong/weak phase transitions are found in the decompactification limit of the ${\mathcal N}=2^*$ theory~\cite{Russo:2013qaa,Russo:2013kea}.

At finite temperature the correspondence can be generalized to the global AdS, with a different boundary manifold $\mathbb{S}\times \mathbb{S}^3$~\cite{Witten:1998zw}. Gauss' law on the compact space forces colored states to disappear, and induces kinematically ``confinement". Early in the 80's the bulk gravity theory is known to undergo a first order phase transition known as the Hawking-Page transition~\cite{Hawking:1982dh}, corresponding to formation of black hole. The free energy of the black hole phase turns to be of order $N^2$, while in the pure thermal phase the temperature dependent part is ${\mathcal O}(N^0)$. This signals the liberation of the colored freedom, or deconfinement. The deconfinement feature could also be seen from another order parameter, the Polyakov loop. In the black hole phase, the Euclidean time circle becomes contractible and the Polyakov loop acquires a nonzero value. The free energy in the deconfined phase
can be expanded in powers in $1/T^2$~\cite{Burgess:1999vb,Zuo:2014vga}, just as proposed in the fuzzy bag model~\cite{Pisarski:2006yk}. Further generalization to the case of a rotating sphere does not change the pattern~\cite{Landsteiner:1999xv}. With a suitable subtraction, one obtains a finite result for the logarithm of the Polyakov loop, which is dominated by a $1/T^2$ term in the whole deconfined phase~\cite{Zuo:2014vga}. Generalization of these results to gauge theory in flat spacetime could help understanding the confinement mechanism. For example, the construction with a dilaton field is not able to generate such corrections consistently~\cite{Zuo:2014iza}.

It will be interesting to perform these calculations in the weak coupling regime. In such a regime perturbation techniques can be employed, and the calculation can be easily generalized to arbitrary gauge theories. The high-temperature expansion for the free energy at weak coupling has actually been derived in ref.~\cite{Burgess:1999vb}, with the Heat-Kernel method~\cite{Schwinger:1951nm,DeWitt1965}. The free energy exhibits a similar expansion in powers of $1/T^2$ as at strong coupling, though with different coefficients. Such an expansion has been recently reproduced with the plane wave matrix model~\cite{Kitazawa:2008mx}. Further calculation in the case of a rotating sphere exhibits similar behavior~\cite{Landsteiner:1999xv}. These results seem to support the speculation that the strong and weak coupling regime could be smoothly interpolated. However, it is not difficult to find out that Gauss' law is never imposed in all these calculations. As a manifestation of this, the resulting free energy is given in a unified expression whether the space manifold is compact or not~\cite{Burgess:1999vb}. To impose Gauss' law properly, one has to derive explicitly the partition function. This is first done for ${\mathcal N}=4$ SYM in ref.~\cite{Sundborg:1999ue}, and generalized to arbitrary gauge theories in ref.~\cite{Aharony:2003sx}. The framework has later been extended to include fundamental matter~\cite{Schnitzer:2004qt}. The phase diagram for ${\mathcal N}=4$ SYM with finite $R$-symmetry chemical potentials is elaborated in~\cite{Yamada:2006rx,Hollowood:2008gp}. The partition function of the system can be nicely expressed as a matrix integral over the holonomy $U$ along the time circle, which can be further simplified as an integral over the eigenvalues of $U$. At low temperature the eigenvalues are distributed uniformly along the unit circle, resulting a vanishing Polyakov loop. In such a phase the density of the gauge-invariant states grows exponentially as the energy, leading to an instability at the Hagedorn temperature~\cite{Hagedorn:1965}. Such an instability induces a transition to a deconfined phase, where the eigenvalues are distributed non-uniformly. The free energy becomes ${\mathcal O}(N^2)$ and the Polyakov loop acquires a nonzero value. A formal analytic solution is given where the eigenvalue distribution is expressed as an infinite sum.
It is suggested that with such a formal solution, one could be able to obtain the high temperature expansion of the free energy and the Polyakov loop~\cite{Aharony:2003sx}. If the interpolation between the strong and weak coupling regime is indeed smooth, one should find similar results as those at strong coupling~\cite{Burgess:1999vb,Zuo:2014vga}. With such an expectation \cite{Zuo:2014oma}, we perform the calculation in this paper.

The paper will be organized as follows. In the next section we will review the matrix model description of free U($N$) gauge theory on a sphere, and derive the high-temperature expansion of the free energy and Polyakov loop. In section \ref{sec.example}  we show the results in pure Yang-Mills theory and ${\mathcal N}=4$ SYM, focusing on the comparison between weak and strong coupling. In the last section we summary our results and give a short discussion.

\section{The free Yang-Mills matrix model and Hagedorn transition}
We consider U($N$) gauge theories on $\mathbb{S}^3$ with only adjoint matter. For convenience the radius of the sphere is set to one. The partition function can be derived either by counting the colorless states or by the Euclidean path integral~\cite{Sundborg:1999ue,Aharony:2003sx}. We follow the formalism in ref.~\cite{Aharony:2003sx} in the whole paper. In the free limit, the partition function can be expressed as a matrix integral over the holonomy $U$ around the time circle as
\begin{equation}
Z(x)=\int [\mathd U] \exp\left\{\sum_{n=1}^\infty \frac{1}{n}\left[z_B(x^n)+(-1)^{n+1}z_F(x^n)\right]\tr (U^n)\tr ((U^\dagger)^n)\right\}, \label{eq.Z}
\end{equation}
where $x\equiv \exp[-1/T]$. $z_B(x)(z_F(x))$ is the bosonic (fermionic) ``single-particle partition function", which encodes the spectrum of oscillators in the corresponding channel,
$z(x)\equiv \sum_i x^{E^i}$.
Explicit expressions for the single-particle partition functions of different spin channels can be found in the appendix of ~\cite{Aharony:2003sx}, which we will discuss later. In the above integral we have neglected the Casimir energy terms, which appear manifestly in the path integral approach.

The stable phase of the theory is determined by the solution of (\ref{eq.Z}) with lowest free energy $F=-T\ln Z$. To find the solutions, we simplify the matrix integral to that over the eigenvalues $\mathe ^{i\theta_i}$ of $U$, with $-\pi<\theta_i\leq \pi$. Then the partition function can be expressed through a pairwise potential as
\begin{equation}
Z(x)=\int_{-\pi}^\pi \mathd \theta_i \cdots \mathd \theta_N ~\mathe ^{-\sum_{i\ne j}~V(\theta_i-\theta_j)},\label{eq.ZV}
\end{equation}
with
\begin{equation}
V(\theta)=-\ln |\sin(\theta/2)|-\sum_{n=1}^\infty \frac{1}{n} [z_B(x^n)+(-1)^{n+1}z_F(x^n)]\cos(n\theta).\label{eq.V-theta}
\end{equation}
The first term, coming from the integration measure, is temperature independent and repulsive. The second term can be shown to be always attractive, and increases from zero to infinite strength as the temperature increases from zero to infinity. As a result, the eigenvalues tend to stay apart at low temperature, and prefer to bunch up at high temperature.

It will be convenient to adapt the method in \cite{Brezin:1977sv} to analyze the large-$N$ limit of the theory. The distribution of the eigenvalue is described by a function $\rho(\theta)$ for $-\pi<\theta\leq \pi$, with the normalization
\begin{equation}
\int _{-\pi}^\pi ~\rho(\theta) ~\mathd \theta =1.\label{eq.rho-norm}
\end{equation}
The effective action $S=-\ln Z$ can then be expressed as
\begin{eqnarray}
S[\rho(\theta)]&=&-\ln Z=N^2\int\mathd \theta~\int \mathd \tilde \theta ~V(\theta-\tilde \theta)~\rho(\theta)~\rho(\tilde \theta)\nonumber\\
&=&\frac{N^2}{2\pi}\sum_{n=1}^\infty |\rho_n|^2 V_n(T),\label{eq.S}
\end{eqnarray}
where $\rho_n$ and $V_n$ are the corresponding Fourier modes of $\rho(\theta)$ and $V(\theta)$ respectively
\begin{eqnarray}
\rho_n&\equiv&\int \mathd \theta~ \rho(\theta) \cos(n\theta)\label{eq.rhon}\\
V_n&\equiv & \int \mathd \theta ~V(\theta) \cos(n\theta)=\frac{2\pi}{n} [1-z_B(x^n)-(-1)^{n+1}z_F(x^n)].
\end{eqnarray}
In the final expression (\ref{eq.S}) a temperature-independent constant term has been subtracted. Since $U$ is the holonomy around the time circle, $\rho_1$ will be the norm of the Polyakov loop $\frac{1}{N}<\tr (U)>$. Later we will see that $\rho_1$ can be considered as the order parameter of the deconfinement phase transition~\cite{Aharony:2003sx}, in the same manner as that in lattice simulation~\cite{Mykkanen:2012ri} and in the gauge/string duality~\cite{Witten:1998zw}.

\subsection{Low temperature behavior and Hagedorn temperature}
Now we are ready to analyze the solutions of minimum action. At low temperature, the single-particle partition function $0<z(x)\ll 1$, and the potential modes are always positive $V_n>0$. The minimum action is achieved with $\rho_n=0$, which represents the uniform distribution of the eigenvalues around the circle. The action for this configuration is zero, and the non-vanishing contribution comes from the fluctuations around such a configuration. This contribution will be $1/N^2$ suppressed compared to (\ref{eq.S}), and of order $N^0$. Explicitly, the contributions from the fluctuation around the configuration $\rho_n=0$ can be integrated to be
\begin{equation}
Z(x)=\prod_{n=1}^\infty\frac{1}{1-z_B(x^n)-(-1)^{n+1}z_F(x^n)}.
\end{equation}
Such a result can also be directly derived by counting the colorless states in the large-$N$ limit~\cite{Sundborg:1999ue,Aharony:2003sx}. The persistence of such a configuration requires $V_n>0$, or $a_n\equiv z_B(x^n)+(-1)^{n+1}z_F(x^n)<1$. Since the single-particle partition function increases monotonously as the temperature increases, the strongest constraint is given by $a_1<1$. If the theory contains at least two oscillating modes, $a_1(x)>1$ when $x\to 1~(T\to \infty)$. Therefore there is a single solution $x=x_H$ of the equation $a_1(x)=1$. As $x\to x_H$ the free energy $F=TS=-T\ln Z$ diverges as
\begin{equation}
F\to T_H\ln (T_H-T).
\end{equation}
Such a divergence is related to the Hagedorn growth of the spectrum density~\cite{Hagedorn:1965,Aharony:2003sx}
\begin{equation}
\rho(E)\propto \mathe ^{ E/T_H}.
\end{equation}
Beyond $x_H$ the potential develops negative modes $V_n<0$, and the eigenvalue distribution of the dominant phase will not be uniform anymore.

\subsection{High temperature behavior and large-$N$ phase transition}
As the temperature increases above $T_H$, the negative modes $V_n$ induce new saddle point of the theory. Since no absolute minimum exists anymore, the minimum action can only be achieved on the boundary of the configuration space, specified by $\rho(\theta)\geq 0$.
Therefore for the new solution above $T_H$, the distribution function must be vanishing in some parts of the circle. For the present system, it turns out that the vanishing area is simply connected. One can therefore assume that $\rho(\theta)\ne0$ only in the interval $[-\theta_0,\theta_0]$. Such a solution at $T_H$ can be immediately constructed~(given later), and shown to have $\theta_0=\pi$. In other words, $\rho(\theta)$ vanishes only at the point $\theta=\pi$. As the temperature goes to infinity, the attrcative part of the potential diverges and compresses the eigenvalues to a single point. As a result, $\rho(\theta)=\delta(\theta)$ and $\theta_0=0$. At a finite temperature $T>T_H$, one needs to solve explicitly the distribution function $\rho(\theta)$ with the corresponding $\theta_0$. For the minimum action configuration, the distribution density function $\rho(\theta)$ satisfies the following equilibrium condition
\begin{equation}
P \int _{-\theta_0}^{\theta_0} \cot \left(\frac{\alpha-\theta}{2}\right)~\rho(\theta)~\mathd \theta=2\sum_{n=1}^{\infty} a_n \rho_n \sin (n\alpha),\label{eq.EoS}
\end{equation}
where $P$ denotes principal value of the integral. Such a condition can be derived directly from the expression (\ref{eq.ZV}) together with the potential (\ref{eq.V-theta}).

\subsubsection{Exact solution above the transition}
The solution of the above equilibrium equation has been given explicitly in~\cite{Aharony:2003sx}, employing the early construction in~\cite{Jurkiewicz:1982iz}.
The Fourier modes $\rho_n$ can be compactly organized into a vector $\vec \rho$ with infinite elements. The definition (\ref{eq.rhon}) and the equilibrium condition (\ref{eq.EoS}) then give two linear constraints on $\vec \rho$, through an infinite matrix $R$ and another vector $\vec A$
\begin{equation}
R \vec \rho=\vec \rho,\quad \vec A\cdot \vec \rho=1. \label{eq.RA}
\end{equation}
In other words, $\vec \rho$ is simply an eigenvector of the matrix $R$ with eigenvalue $1$, and further normalized so that its dot product with $\vec A$ is $1$. Elements of $R$ and $\vec A$ are polynomials of $s^2\equiv \sin (\theta_0/2)$, with the coefficients linear in $a_n$. The vector $\vec A$ is defined through the Legendre polynomials $P_n$ as
\begin{equation}
A_n\equiv a_n(P_{n-1}(1-2s^2)-P_n(1-2s^2)).\label{eq.A}
\end{equation}
The matrix $R$ is given in a similar way
\begin{equation}
R_{nl}\equiv a_l\sum_{k=1}^l(B^{n+k-1/2}(s^2)+B^{|n-k+1/2|}(s^2))P_{l-k}(1-2s^2),\label{eq.R}
\end{equation}
where the functions $B^{n-\frac{1}{2}}(s^2)$ is defined as
\begin{equation}
B^{n-\frac{1}{2}}(s^2)\equiv \frac{1}{\pi} \int _{-\theta_0}^{\theta_0}\sqrt{\sin^2(\frac{\theta_0}{2})-\sin^2(\frac{\theta}{2})}\cos((n-\frac{1}{2})\theta)~\mathd \theta.
\end{equation}
(\ref{eq.RA}) immediately leads to $\det (1-R)=0$, which determines $\theta_0$ in terms of all $a_n$. Replacing the first row of the matrix $1-R$ by $\vec A$, one obtains a new matrix $M$. With the constraint $\det (1-R)=0$, (\ref{eq.RA}) is solved by
\begin{equation}
\vec \rho = M^{-1} e_1, \label{eq.rho-M}
\end{equation}
where $e_1=(1,0,0,\cdots)$. From $\theta_0$ and $\rho_n$ one can then recover the function $\rho(\theta)$
\begin{equation}
\rho(\theta)= \frac{1}{\pi} \int _{-\theta_0}^{\theta_0}\sqrt{\sin^2(\frac{\theta_0}{2})-\sin^2(\frac{\theta}{2})}\sum_{n=1}^\infty Q_n~\cos((n-\frac{1}{2})\theta),\label{eq.rho-theta}
\end{equation}
with $Q_n$ defined as
\begin{equation}
Q_n\equiv 2\sum _{l=0}^\infty a_{n+l}\rho_{n+l}P_l(\cos(\theta_0)).
\end{equation}
One can check that such a distribution function indeed satisfies the equilibrium condition (\ref{eq.EoS}). Integrating the potential with such a distribution, one finally obtains the effective action and the free energy.

\subsubsection{Perturbative expansion above phase transition}
With the formal solution (\ref{eq.rho-theta}) we can go on to study the behavior near the phase transition.
At the transition, $V_1=0$ and  $\rho_1$ acquires a nonzero value, while $\rho_{n>1}=0$. The eigenvalue distribution is given by
\begin{equation}
\rho(\theta)=\frac{1}{2\pi}+ \frac{\rho_1}{\pi} \cos(\theta).\label{eq.rho-TH}
\end{equation}
When $T>T_H$ the minimum action occurs at the boundary of the configuration space $\rho(\theta)
\ge0$ and the distribution therefore vanishes for some part of the circle. Due to the convex property of the boundary region, the minimum action configuration could be continuously extended to the phase transition point. Thus at $T_H$ the vanishing segment shrinks to a point. This fixes $\rho_1=1/2$. A similar eigenvalue distribution was found in the large-$N$ strong/weak transition of 2D U($N$) lattice gauge theory~\cite{Gross:1980he,Wadia:2012fr}. For SU($N$) pure gauge theory in 4 dimensional flat spacetime, lattice data shows that the renormalized Polyakov loop at the critical temperature is indeed close to $1/2$, for $N=3$~\cite{Kaczmarek:2002mc,Gupta:2007ax} and $N=4, 5$~\cite{Mykkanen:2012ri}. The specific distribution of the eigenvalues may provide an explanation of such an observation.

The above discussion on the Polyakov loop can be deduced explicitly from the matrix solution given in the previous subsection.
When $T\to T_H^+$, $\rho_{n>1}\to 0$ and the distribution is given by (\ref{eq.rho-TH}). Such a distribution can at most vanish at a single point, leading to $\theta_0=\pi$.
Then it is not difficult to check that at $T_H$,
\begin{equation}
R=diag(a_1,a_2,a_3,\cdots).
\end{equation}
Therefore, the nonzero elements of the matrix $M$ are in the first row or the diagonal line. The first row of $M$ is simply $\vec A$, with
\begin{equation}
M_{11}=A_1=2a_1 s^2.
\end{equation}
The first moment $\rho_1=(M^{-1})_{11}=1/(2 a_1 s^2)$. At the phase transition, $a_1(x_H)=1$ and $\theta_0=\pi$, giving $\rho_1=1/2$.

The resulting minimum action slightly above $T_H$ is
\begin{eqnarray}
S_{min}&=&\frac{N^2}{8\pi} V_1'(T_H)(T-T_H)+\cdots \nonumber\\
       &=&-\frac{N^2}{4} \frac{x_H}{T_H^2} a_1'(x_H) (T-T_H)+\cdots
\end{eqnarray}
where``$\cdots$" denotes suppressed terms as $T\to T_H^+$. And then for the free energy
\begin{equation}
\lim_{N\to \infty} \frac{1}{N^2} F_{T\to T_H^+} \approx -\frac{1}{4}\frac{a_1'(x_H)x_H}{T_H}(T-T_H).
\end{equation}
While for $T<T_H$, $\lim_{N\to \infty} \frac{1}{N^2} F(T)=0$. So in the large-$N$ limit a first order phase transition occurs at $T_H$, with $F/N^2$ and $\rho_1$ as the order parameters. Since the degrees of freedom increase from ${\mathcal O}(1)$ to ${\mathcal O}(N^2)$ and the Polyakov loop acquires a nonzero value, it corresponds to the deconfinement transition. The transition order may change when a nonzero coupling is turned on. Depending on the coefficient of the quartic term of $\rho_1$, the transition order will be different~\cite{Aharony:2003sx}.

\subsubsection{Asymptotic expansion at high temperature}
We can further study the high-temperature expansion of the free energy and the Polyakov loop with the formal solution (\ref{eq.rho-theta}). To do this we first need the detailed expressions of the one-particle partition functions. For vectors, scalars and chiral fermions in the adjoint representation, the corresponding partition functions on $\mathbb{S}\times \mathbb{S}^3$ are given by
\begin{equation}
z_{V4}(x)=\frac{6x^2-2x^3}{(1-x)^3},\quad z_{S4}(x)=\frac{x+x^2}{(1-x)^3}, \quad z_{F4}(x)=\frac{4 x^{3/2}}{(1-x)^3}.
\end{equation}
Here conformal coupling of the scalars has been assumed. Their high-temperature expansions are
\begin{eqnarray}
z_{V4}(x)&\to& 4 T^3 - 2 T + 1 - \frac{11}{60 T}+{\mathcal O}(\frac{1}{T^3}),\label{eq.zV}\\
z_{S4}(x)&\to& 2 T^3 - \frac{1}{120 T}+{\mathcal O}(\frac{1}{T^3}),\label{eq.zS}\\
z_{F4}(x)&\to& 4 T^3 - \frac{T}{2} + \frac{17}{480 T}+{\mathcal O}(\frac{1}{T^3})\label{eq.zF}.
\end{eqnarray}
All the fields has the leading behavior
\begin{equation}
z(x) \to 2 \mathbb{N}^{dof}~T^3,\label{eq.z-asym}
\end{equation}
with $\mathbb{N}^{dof}$ the number of single-particle degrees of freedom. The sub-leading term of ${\mathcal O}(T)$ does not appear in the scalar excitation due to the conformal coupling.
Finally, the terms of ${\mathcal O}(1/T)$ simply reflect the effects of the Casimir energy. If we keep the Casimir energy at the very beginning in the path integral formalism~\cite{Aharony:2003sx}, these terms will be canceled. For a specific theory containing $n_S$ scalars and $n_F$ spinors, we need to sum them into
\begin{equation}
z_B(x)=z_{V4}(x)+n_S z_{S4}(x),~~~\quad z_F(x)=n_F z_{F4}(x).
\end{equation}

The free energy depends on both the single-particle partition functions and the eigenvalue distribution $\rho(\theta)$.
When $T\to \infty$, the attractive part of the potential (\ref{eq.V-theta}) increases to infinite strength. Accordingly, the minimum action distribution $\rho(\theta)\to \delta(\theta)$ and $\rho_n\to 1$.
Let us first take the approximation $\rho_n=1$ and neglect the deviations at finite temperature. Then the free energy is completely determined by the single-particle partition functions, and has the following expansion
\begin{eqnarray}
\frac{F}{N^2}&=&-(4+2 n_S+\frac{7}{2}n_F)~\zeta(4)~T^4+(2+\frac{1}{4}n_F)~\zeta(2)~T^2\nonumber\\
&& +(\frac{11}{360}+\frac{1}{120} n_S+\frac{17}{480} n_F)~\zeta(0)+{\mathcal O}(\frac{1}{T^2}).
\label{eq.FE-HT}
\end{eqnarray}
Here we use the Riemann zeta function to regularize the apparently divergent summation.
One finds that it is a power series in $1/T^2$, in exactly the pattern proposed in ref.~\cite{Pisarski:2006yk}.
It will be instructive to compare the above expansion with that obtained with the Heat-Kernel method~\cite{Burgess:1999vb}. In such an approach, the free energy is expressed through the coefficients $\alpha_k$ of the derivative expansion
\begin{equation}
F=-\frac{1}{16}\sum_{k=0}^\infty  \left[\tr _V \alpha_k+\tr _S \alpha_k+(1-2^{2k-3})\tr _F \alpha_k\right]\,\Delta_k,\label{eq.FE-HK}
\end{equation}
with $\Delta_k=2(4T^2)^{2-k}\Gamma(2-k)\zeta(4-2k)$. The specific form of such an expansion can be traced back to the Heat-Kernel equation and the analyticity of the derivative expansion. For U($N$) gauge theories on a three-space of constant curvature $\kappa$, the trace of the first few coefficients $\alpha_k$ in the adjoint representation are~\cite{Burgess:1999vb}
\begin{eqnarray}
&&\tr_V \alpha_0=2 N^2,\quad \tr_S \alpha_0 =n_S N^2,\quad \tr _F \alpha_0 = 2n_F N^2,\nonumber\\
&& \tr_V \alpha_1=\frac{2}{3} N^2 R,\quad \tr_S \alpha_1 =0,\quad \tr _F \alpha_1 = \frac{1}{6}n_F N^2 R ,\nonumber\\
&& \tr_V \alpha_2=\tr_S \alpha_2=\tr_F \alpha_2=0,
\end{eqnarray}
with the Ricci scalar $R=-6 \kappa$. Again the scalars are assumed to be conformally coupled. With these coefficients, the free energy can be expressed as
\begin{equation}
\frac{F}{N^2}=-(4+2 n_S+\frac{7}{2}n_F)~\zeta(4)~T^4+(2+\frac{1}{4}n_F)~\zeta(2)~\kappa T^2+{\mathcal O}(\frac{1}{T^2}).\label{eq.FE-HK2}
\end{equation}
For a sphere of unit radius $\kappa=1$, we reproduce exactly the expansion (\ref{eq.FE-HT}) up to the constant Casimir term. $\kappa=0$ and $\kappa=-1$ correspond to the flat and hyperbolic spaces respectively. However, eq.~(\ref{eq.FE-HK2}) can not be valid equally well for both the compact and the incompact cases. As illustrated clearly in ref.~\cite{Witten:1998zw}, when the space manifold is compact additional constraints appear due to Gauss' law. Such constraints are imposed in the partition function (\ref{eq.Z}) through the integration over the group characters. They will be completely relaxed when we take the approximation $\rho_n= 1$. From the Heat-Kernel derivation it is clear that no additional constraint has been imposed for the special case $\kappa=1$. As a result, the two expressions from the two approaches coincide with each other.

In the present formalism the approximation $\rho_n=1$ can not be exact since we have $\rho_1=1/2$ and $\rho_{n>1}=0$ at the phase transition. At large but finite temperature, the eigenvalues will be distributed in the interval $[-\theta_0, \theta_0]$ with $\theta_0\ne 0$. Since $s^2\equiv\sin^2 (\theta_0/2)$ is small in the asymptotic region, we first keep only linear terms of $s^2$ in $\vec A$ and $R$
\begin{equation}
A_k=2ka_k (s^2+{\mathcal O}(s^4)),\quad R_{nk}=2ka_k (s^2+{\mathcal O}(s^4)).\label{eq.R-HT}
\end{equation}
The constraint $\det (1-R)=0$ becomes
\begin{equation}
\det (1-R)=1-\sum_{n=1}^\infty 2n a_n (s^2+{\mathcal O}(s^4))=0.\label{eq.dR-HT}
\end{equation}
From this one obtains
\begin{eqnarray}
s^2&=&1/\sum_{n=1}^\infty 2n a_n+{\mathcal O}(s^4)\nonumber\\
&=&\frac{1}{4\zeta(2)(2+n_S+n_F)}\frac{1}{T^3}-\frac{\zeta(0)(4-n_F)}{16\zeta(2)^2(2+n_S+n_F)^2}\frac{1}{T^5}+{\mathcal O}(\frac{1}{T^6}),\label{eq.s2-HT}
\end{eqnarray}
where we have again used the function $\zeta(x)$ to regularize the sub-leading term.
With the asymptotic expansion of $s^2$ we can go ahead to obtain the Fourier modes $\rho_n$. From the asymptotic matrix elements (\ref{eq.R-HT}) it is not difficult to find the modes $\rho_n$ are given by
\begin{eqnarray}
\rho_1&=&\frac{1-\sum_{n=2}^\infty 2n a_n s^2}{2a_1 s^2}+{\mathcal O} (s^2),\nonumber\\
\rho_{n>1}&=&1+{\mathcal O} (s^2).
\end{eqnarray}
With the constraint (\ref{eq.dR-HT}) and the asymptotic solution (\ref{eq.s2-HT}) one concludes
\begin{equation}
\rho_n=1+{\mathcal O} (s^2)=1+{\mathcal O}( T^{-3}).
\end{equation}

In order to obtain explicitly the subleading terms in $\rho_n$, we have to derive the higher power terms in $\vec A$ and $R$.
Expanding (\ref{eq.A}) and (\ref{eq.R}) up to terms of $s^4$, one finds
\begin{equation}
A_l= 2la_l s^2+\delta  A_l, \quad R_{ml}=2la_l s^2+\delta R_{ml},
\end{equation}
with
\begin{eqnarray}
{\delta A}_{l}&=&a_l\big[-l(l^2-1)  s^4+{\mathcal O} (s^6)\big],\nonumber\\
\delta R_{ml}&=&a_l\Bigg\{-\frac{s^4}{2}\sum_{k=1}^l \bigg[(m+k-1)(m+k)\nonumber\\
&&~~~~~~~~~~~~~~~~+(|m-k+\frac{1}{2}|-\frac{1}{2})(|m-k+\frac{1}{2}|+\frac{1}{2}) \nonumber\\
&&~~~~~~~~~~~~~~~~+4(l-k)(l-k+1))\bigg]+{\mathcal O} (s^6)\Bigg\}.
\end{eqnarray}
With these we can find the corresponding expansions for the matrix $M$ by replacing the first row of $1-R$ by $\vec A$. Some efforts are needed to derive the inverse of $M$.
The expressions for $M^{-1}$ are a little complicated and we do not list them explicitly. $M^{-1}$ simplifies a lot when the constraint $\det (1-R)=0$ at the same order is substituted. According to (\ref{eq.rho-M}) the first column of $M^{-1}$ then gives
\begin{eqnarray}
\rho_n&=& (M^{-1})_{n1}\nonumber\\
&=&1+\sum_{k=1}^\infty (\delta R_{nk}-\delta A_k)+{\mathcal O} (s^4)\nonumber\\
&=& 1-\frac{n^2}{2} \sum_{k=1}^\infty a_k \left[2k  s^4+{\mathcal O} (s^6)\right]\nonumber\\
&=& 1-\frac{n^2}{2} s^2+{\mathcal O} (s^4).\label{eq.rho-s}
\end{eqnarray}
In obtaining the last expression we have used the constraint (\ref{eq.dR-HT}). Substituting the expansion (\ref{eq.s2-HT}) for $s^2$, one can reexpress the expansion in temperature as
\begin{eqnarray}
\rho_n&=&1-\frac{n^2}{2}\frac{1}{\sum_{k=1}^\infty2ka_k}+{\mathcal O} (s^4)\nonumber\\
&=&1-\frac{n^2}{8\zeta(2)(2+n_S+n_F)}\frac{1}{T^3}+\frac{n^2\zeta(0)(4-n_F)}{32\zeta(2)^2(2+n_S+n_F)^2}\frac{1}{T^5}+{\mathcal O}\left(\frac{1}{T^6}\right).\label{eq.rho-T}
\end{eqnarray}
In particular, from the derivation it is clear that the $1/T^3$ terms of $\rho_n$ are completely determined by the asymptotic terms of all $a_k$. As we show before all the single-particle partition functions share the same asymptotic behavior (\ref{eq.z-asym}), which is guaranteed by scale invariance. Therefore the leading power correction of $\rho_n$ is the same in different theories, though the coefficients differ due to different field contents.

Such deviations will then modify the previous expansion for the free energy (\ref{eq.FE-HT}). To obtain this, it will be convenient to rewrite the minimal effective action as
\begin{equation}
\frac{S^{\rm{min}}_{T>T_H}}{N^2}=\sum_{n=1}^\infty  \frac{|\rho_n|^2}{n} \left[1-a_n\right]= \sum_{n=1}^\infty  \frac{|\rho_n|}{n} \left[1-a_n\right].\label{eq.Smin}
\end{equation}
The derivation of the final expression is not quite obvious. One has to transform the corresponding 2d integral to one dimensional, and then use the stability condition (\ref{eq.EoS}).
Inputting the asymptotic expansion (\ref{eq.rho-s}) one finally obtains
\begin{equation}
\frac{S^{\rm{min}}_{T>T_H}}{N^2}= \sum_{n=1}^\infty  \frac{1}{n} \left[1-a_n\right]+\frac{1}{4}+{\mathcal O}(s^2).
\end{equation}
The first term gives rise to the expansion (\ref{eq.FE-HT}) exactly. And the ${\mathcal O}(s^2)$ corrections to the moments $\rho_n$ combine to give a universal constant term in the effective action. As a result, the expansion (\ref{eq.FE-HT}) of the free energy is modified as
\begin{eqnarray}
\frac{F}{N^2}&=&-(4+2 n_S+\frac{7}{2}n_F)~\zeta(4)~T^4+(2+\frac{1}{4}n_F)~\zeta(2)~T^2\nonumber\\
&& +\frac{1}{4}T+(\frac{11}{360}+\frac{1}{120} n_S+\frac{17}{480} n_F)~\zeta(0)+{\mathcal O}(\frac{1}{T^2}).
\label{eq.FE-HT2}
\end{eqnarray}
Therefore the expansion pattern in powers of $1/T^2$ is broken by a linear term in $T$, which is universal among different theories. For massless scalar fields on a curved spacetime, such a term often results from non-local effects and does not show up in the derivative expansion~\cite{Gusev:1998rp}. Moreover, no logarithmic terms as speculated in \cite{Burgess:1999vb} appear. This is due to the cancelation of  the ``1" in the expansion (\ref{eq.zV}) with that from the repulsive potential. We will later find that when we take the truncated approximation $a_{n>k}=0$, such a cancelation is spoiled. A logarithmic term in temperature will appear, simply from the infinite summation of the repulsive contributions.



\subsubsection{Approximate truncated solutions}
The full solution with the infinite matrix $R$ is not easy to obtain. However, qualitative properties can be found by truncating the infinite matrix to a finite one. This can be achieved by setting $a_{n>k}=0$. Since the contribution from $a_n$ with large $n$ is power suppressed, such a truncation should give an arbitrarily good approximation as long as $k$ is large enough. However, one must also be very careful for such an approximation may also bring about artifacts.

With the truncation $a_{n>k}=0$, the minimal action (\ref{eq.Smin}) simplifies as
\begin{eqnarray}
\frac{S^{\rm{min}}_{\rm{trunc}}}{N^2}&=& -\sum_{n=1}^k  \frac{|\rho_n|}{n} a_n + \sum_{n=1}^\infty  \frac{|\rho_n|}{n} \nonumber\\
&=& -\sum_{n=1}^k  \frac{|\rho_n|}{n} a_n +\frac{1}{2} \int_{-\theta_0}^{\theta_0}~\sigma(\theta) \cot (\frac{\theta}{2}) \mathd \theta -\frac{1}{2}\ln s^2 -\ln 2,\label{eq.trunc}
\end{eqnarray}
where $\sigma(\theta)$ is the antiderivative of $\rho(\theta)$, $\sigma(\theta)\equiv \int _0^ \theta ~\rho(\theta) ~\mathd \theta$.
In deriving the above expression we have used again the stability condition (\ref{eq.EoS}). The logarithmic term in $s^2$, and thus in temperature, shows up as expected from the infinite summation over the repulsive contributions. It appears at a higher power in $T$ than the possible logarithmic term due to infrared divergence~\cite{Burgess:1999vb}. This also indicates that it could be artificial. The distribution function also simplifies since $Q_{n>k}=0$
\begin{equation}
\rho(\theta)= \frac{1}{\pi} \int _{-\theta_0}^{\theta_0}\sqrt{\sin^2(\frac{\theta_0}{2})-\sin^2(\frac{\theta}{2})}\sum_{n=1}^k Q_n~\cos((n-\frac{1}{2})\theta).\label{eq.rho-theta2-trunc}
\end{equation}
With finite terms in the density function $\rho(\theta)$, the integral in (\ref{eq.trunc}) is regular in the whole region. The integral can be done term by term with the truncated expansion (\ref{eq.rho-theta2-trunc}), resulting a polynomial in $s^2$ which remains finite as $T\to \infty$.
Taking $k$ larger and larger, one can then approximate the exact distribution with higher and higher accuracy.

The simplest case is $k=1$, for which the solution has been studied a lot. In such an approximation, the constraint $\det (1-R)=0$ simplifies to
\begin{equation}
0=\det (R_{1\times1}-1)=a_1(2s^2-s^4)-1,\label{eq.R11}
\end{equation}
giving
\begin{equation}
s^2=1-\sqrt{1-1/a_1}.
\end{equation}
The vector $\vec A$ has one nonzero element $A_1=2a_1 s^2$, leading to the Polyakov loop
\begin{equation}
\rho_1=(2a_1 s^2)^{-1}=1-s^2/2.\label{eq.rho1-trunc}
\end{equation}
 Note that this is in accordance with the general expansion (\ref{eq.rho-s}). $\rho_1$ in turn completely determines the eigenvalue distribution
\begin{equation}
\rho(\theta)=\frac{1}{\pi \sin^2(\frac{\theta_0}{2})}\sqrt{\sin^2(\frac{\theta_0}{2})-\sin^2(\frac{\theta}{2})}\cos(\frac{\theta}{2}).\label{eq.rho-trunc}
\end{equation}
With such a distribution the integral in (\ref{eq.trunc}) can be carried out immediately, giving
\begin{equation}
\frac{1}{2} \int_{-\theta_0}^{\theta_0}~\sigma(\theta) \cot (\frac{\theta}{2}) \mathd \theta=\frac{1}{2}+\ln 2.
\end{equation}
Combining all the pieces, the effective action is simply
\begin{equation}
\frac{S^{\rm{min}}_{\rm{trunc}}}{N^2}=-\left(\frac{1}{2s^2}+\frac{1}{2}\ln s^2 -\frac{1}{2}\right).
\end{equation}
It is not difficult to check that when $T\to T_H^+$,
\begin{equation}
\rho_1\to \frac{1}{2},\quad \frac{F_{\rm{trunc}}}{N^2}\sim -\frac{1}{4}\frac{a_1'(x_H)x_H}{T_H}(T-T_H).
\end{equation}
 We will use this truncated solution in the next section to show the qualitative properties in different theories. One can go further to obtain the solutions with larger $k$. Since the structure is essentially similar, we do not show the details anymore. For the case of $k=2$, one could find some results in ref.~\cite{Jurkiewicz:1982iz}.

\section{Two specific theories}\label{sec.example}
Now we give the explicit results in two specific gauge theories, extending the discussion in~\cite{Aharony:2003sx} to the high temperature region.
\subsection{Pure Yang-Mills theory}
For pure Yang-Mills theory we have only the gauge field.
The Hagedorn transition occurs when
\begin{equation}
a_1= z_{V4}(x)=1.
\end{equation}
So $x_H=2-\sqrt{3}$, $T_H=-1/\ln(2-\sqrt{3})\simeq 0.76$. In order to make qualitative comparison with the lattice data, we show explicitly the results for the Polyakov loop $\rho_1$ and the free energy density,
\begin{equation}
f=F/V_{S^3},\quad V_{S^3}=2\pi^2.
\end{equation}
According to (\ref{eq.rho-T}) and (\ref{eq.FE-HT2}), $\rho_1$ and $f$ have the following high-temperature expansions:
\begin{eqnarray}\label{eq.YM-as}
\rho_1 &=&1-\frac{3}{8\pi^2}\frac{1}{T^3}-\frac{9}{16\pi^4}\frac{1}{T^5}+{\mathcal O}\left(\frac{1}{T^6}\right)\label{eq.rho-YM}\\
\frac{f}{N^2}&=&-\frac{\pi^2}{45}T^4\left[1-\frac{15}{2\pi^2}\frac{1}{T^2}-\frac{45}{8\pi^4}\frac{1}{T^3}+\frac{99}{288\pi^4}\frac{1}{T^4}+{\mathcal O}\left(\frac{1}{T^6}\right)\right]\nonumber\\
&\approx&-\frac{\pi^2}{45}T^4\left[1-1.3~\frac{T_H^2}{T^2}-0.13~\frac{T_H^3}{T^3}+0.01~\frac{T_H^4}{T^4}+{\mathcal O}\left(\frac{1}{T^6}\right)\right].\label{eq.f-YM}
\end{eqnarray}

If the radius of the sphere is taken to infinity, we recover the theory in flat spacetime. The phase transition in infinite volume is believed to occur in the strong coupling regime and not easy to study. However, by formulating the theory on a discrete lattice and doing numerical simulation, the Polyakov loop and thermodynamics have been obtained in great detail. The lattice data  show that, for pure SU($N$) gauge theory with various $N$, both the Polyakov loop and the free energy density acquire mainly quadratic corrections in temperature~\cite{Megias:2005ve,Pisarski:2006yk,Panero:2009tv,Mykkanen:2012ri}. The Polyakov loop between $T_c$ and a few times $T_c$ is well fitted by the formula
\begin{equation}
- 2 \log {\mathcal L} = a + b \left(\frac{T_c}T \right)^2,\label{eq.rho-L}
\end{equation}
with~the parameters slightly dependent on $N$~\cite{Megias:2005ve,Megias:2007pq,Mykkanen:2012ri}
\begin{equation}
a\sim -(0.1-0.3),~~~~b\sim 1.1-1.8.
\end{equation}
The free energy density in roughly the same region can also be well fitted as~\cite{Pisarski:2006yk,Panero:2009tv}
\begin{equation}
\frac{f}{N^2}=-\frac{\pi^2}{45}T^4\left[1-\frac{f_3}{2}~\frac{T_c^2}{T^2}+\frac{f_4}{2}~\frac{T_c^4}{T^4}\right],
\end{equation}
and the fitted parameters extrapolated to large $N$ have the central values $f_3\sim 1.8,~f_4\sim -0.2$.

\begin{figure}[ht]
\centering
	\includegraphics[width=\textwidth]{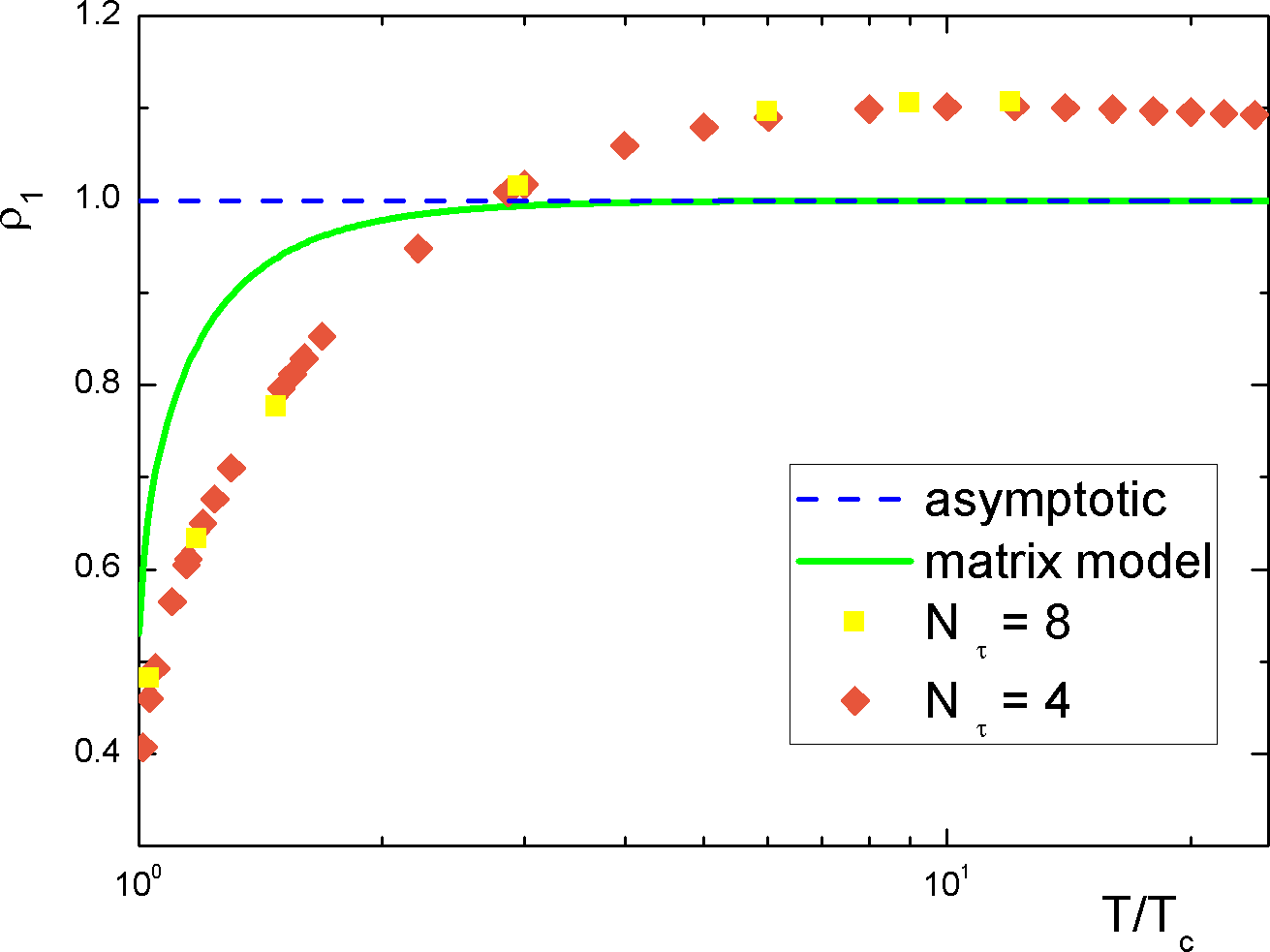}
\caption{\it Matrix model result in the truncated approximation for the Polyakov loop of pure Yang-Mills theory on a sphere, in comparison to the lattice data for pure SU($3$) gauge theory in flat spacetime~\cite{Gupta:2007ax}. The asymptotic value $\rho_1=1$ is also plotted.}\label{fig:Polyakov-YM}
\end{figure}

Comparing the results from the two approaches in the region close to the phase transition, we can find the difference between weak and strong coupling, or between small finite and infinite volume. The specific expansion of the free energy density in powers of $1/T^2$ is broken in the compact case, due to the universal linear term in $T$. One may argue that such a term will have little numerical effect and be buried in the numerical fit. Indeed the numerical coefficient of the $1/T^3$ term in (\ref{eq.f-YM}) is much smaller than the leading ones. Nevertheless, the difference in the Polyakov loop is quite significant. The expansion pattern is completely different. The deviation is given mainly by odd powers of inverse temperature in (\ref{eq.rho-YM}), while in (\ref{eq.rho-L}) it is in ever powers only. Notice that this difference is not just conceptual, it actually has significant numerical effect. Due to missing of quadratic correction in (\ref{eq.rho-YM}), the total contributions of corrections up to ${\mathcal O}(1/T^5)$ decrease the Polyakov loop by only $10\%$ at the phase transition. In other words, a long power series combine to give the special value $\rho_1=1/2$ at the phase transition. Away from $T_H$, most of the power terms are strongly suppressed, increasing $\rho_1$ quickly to the asymptotic value. In contrast, the situation in the formula (\ref{eq.rho-L}) is quite different. Here a single quadratic correction decreases the Polyakov loop to a value around $1/2$ at $T_c$. The logarithm of the Polyakov loop is dominated by the quadratic correction up to several $T_c$, until finally taken over by the perturbative contributions. Such a difference is shown clearly in  FIG.~\ref{fig:Polyakov-YM}. Since we do not have the exact solution of $\rho_1$ in the full theory, we show in FIG.~\ref{fig:Polyakov-YM} the qualitative behavior using the truncated solution~(\ref{eq.rho1-trunc}).

Similar behavior is also found in the effective matrix models~\cite{Meisinger:2001cq,Dumitru:2010mj,Dumitru:2012fw}. Specifically, the deviation of the Polyakov loop from one is suppressed by powers higher than $1/T^2$. In the present theory the deviation is always $1/T^3$, while in ~\cite{Dumitru:2010mj,Dumitru:2012fw}, it is estimated to be of $1/T^4$.  Moreover, in our derivation it is clear that the $1/T^3$ correction in the free energy density is induced from the corresponding terms in $\rho_n$. Therefore if one wants to eliminate these odd thermal corrections, the Polyakov loop will be forced to approach one quickly, with the deviation visible only near the transition temperature. This is just the observation in the effective matrix models~\cite{Dumitru:2010mj,Dumitru:2012fw}.

\subsection{${\mathcal N}=4$ SYM}
The results in ${\mathcal N}=4$ SYM is quite similar. We simply list all the results, in order to make comparison with the corresponding ones at strong coupling~\cite{Burgess:1999vb,Zuo:2014vga}. The field content of this theory is specified by
\begin{equation}
n_S=6,\quad n_F=4.
\end{equation}
The Hagedorn temperature is $x_H=7-4\sqrt{3}$ and $T_H=-1/\log (7-4\sqrt{3})\simeq 0.38$, which is exactly one half of that in the pure Yang-Mills. The first few terms in the Polyakov loop $\rho_1$ and the free energy density are
\begin{eqnarray}
\rho_1&=&1-\frac{1}{16\pi^2}\frac{1}{T^3}+{\mathcal O}\left(\frac{1}{T^6}\right)\label{eq.l-SYM1}\\
\frac{f}{N^2}&=&- \frac{\pi^2 T^4}{6} + \frac{T^2}{4}+\frac{T}{8\pi^2} -\frac{1}{18\pi^2}+{\mathcal O}\left(\frac{1}{T^2}\right).\label{eq.f-SYM1}
\end{eqnarray}
The ${\mathcal O} (1/T^5)$ term in $\rho_1$ vanishes due to the cancelation between the vector and fermionic parts.


We can proceed to compare the results with those at strong coupling, obtained through the gauge/gravity duality~\cite{Witten:1998zw,Burgess:1999vb,Zuo:2014vga}. The theory exhibits a first order phase transition, the so-called Hawking-Page transition, due to formation of black hole in the global AdS~\cite{Hawking:1982dh,Witten:1998zw}.
The free energy density above the transition is expressed through the black hole horizon $r_+$ as
\begin{equation}
f=-\frac{N^2}{8\pi^2}r_+^2(r_+^2-1),
\end{equation}
with
\begin{equation}
r_+=\frac{\pi}{2}(T+\sqrt{T^2-T_{\rm{min}}^2}).
\end{equation}
Here $T_{\rm{min}}=\sqrt{2}/\pi$ is the lowest temperature for the black hole solutions to exist. The transition occurs slightly above $T_{\rm{min}}$
\begin{equation}
T_{HP}=\frac{3}{2\pi}\sim 0.48,
\end{equation}
which is also slightly larger than the Hagedorn temperature $T_H$.
At high temperature $f$ can be expanded as \cite{Burgess:1999vb,Zuo:2014vga}
\begin{equation}
\frac{f}{N^2}=- \frac{\pi^2 T^4}{8} + \frac{3T^2}{8} -\frac{3}{16\pi^2} +{\mathcal O} \left(\frac{1}{T^2}\right).\label{eq.f-SYM2}
\end{equation}
Generalized expressions for $\kappa=-1$ have also been given in ref.~\cite{Burgess:1999vb}. As well known, the leading term gives the exact result in flat spacetime, which differs from the weak coupling result (\ref{eq.f-SYM1}) by a factor of $3/4$~\cite{Gubser:1996de}. The correction to the coefficient of the leading term has been calculated both at strong coupling~\cite{Gubser:1998nz} and at weak coupling~\cite{Fotopoulos:1998es}. At strong coupling the correction is positive, while that at weak coupling is negative. These corrections indicate that the free energy of the flat theory could be smoothly interpolated between the strong and weak coupling limits. Such a smooth interpolation is argued to be valid also on the sphere, based on the similarity between (\ref{eq.FE-HK2}) and (\ref{eq.f-SYM2})~\cite{Burgess:1999vb}.
However, as we discussed in the previous section, the derivative expansion (\ref{eq.FE-HK}), and accordingly (\ref{eq.FE-HK2}), must be modified due to the additional constraints on the compact space manifold. The correct high temperature expansion of the free energy density receives a universal term in $T$ as given in (\ref{eq.f-SYM1}). Higher odd powers of $1/T$ are also expected to appear in the expansion (\ref{eq.f-SYM1}), which then differs further from the expansion pattern in (\ref{eq.f-SYM2}).

\begin{figure}[ht]
\centering
	\includegraphics[width=\textwidth]{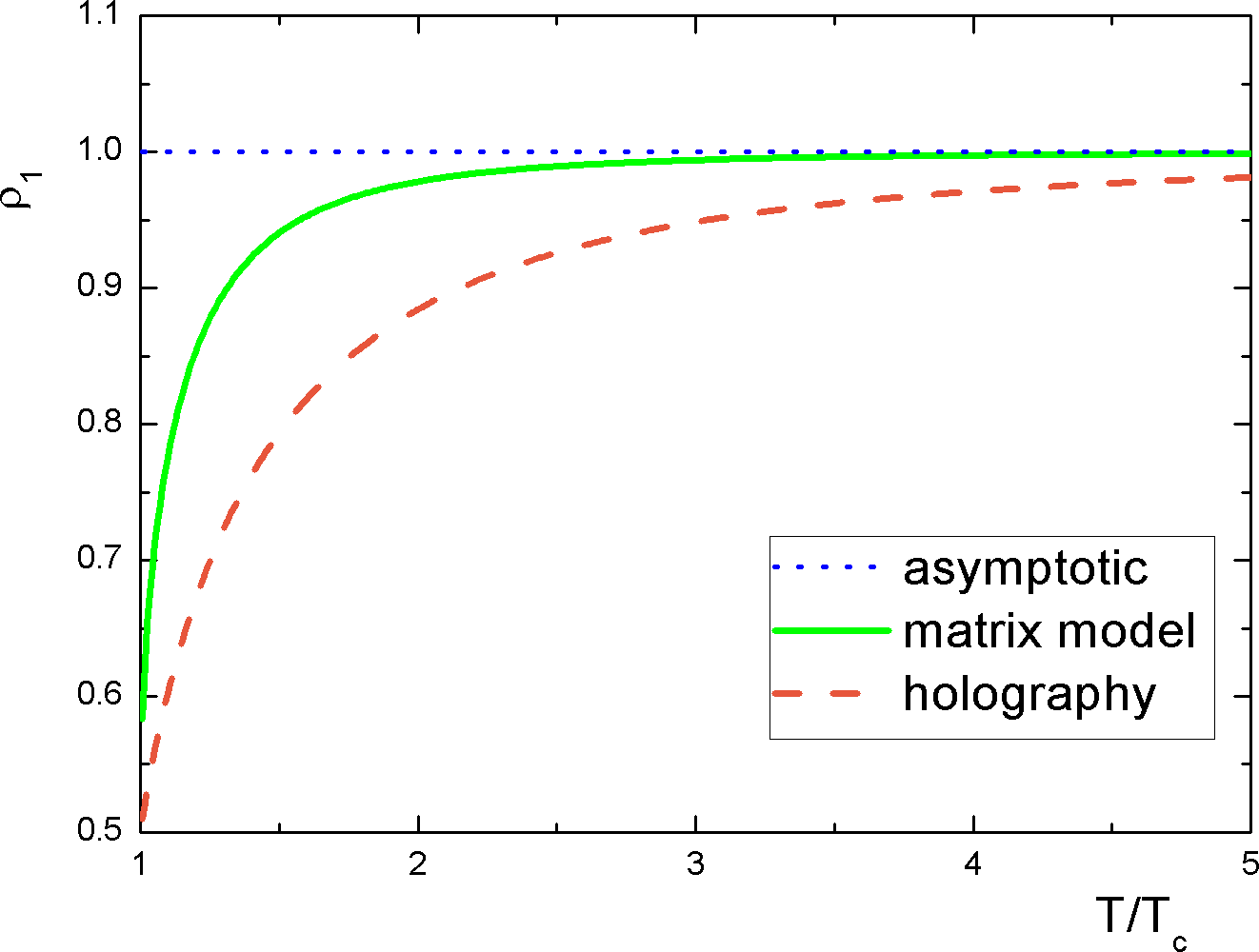}
\caption{\it Matrix model result in the truncated approximation for the Polyakov loop of ${\mathcal N}=4$ SYM on a sphere, in comparison to the holographic result. The asymptotic value $\rho_1=1$ is also plotted.}\label{fig:Polyakov-SYM}
\end{figure}

One could go on to check if such a difference exists also for the Polyakov loop. From the gauge/string duality, the Polyakov loop can be derived from the minimal area of the string worldsheet ending on the loop~\cite{Maldacena:1998im,Rey:1998ik}. In the black hole phase the loop is contractible and the Polyakov loop takes a nonzero value~\cite{Witten:1998zw}. With a proper subtraction, it is given by the following expression~\cite{Zuo:2014vga}
 \begin{eqnarray}
 {\mathcal L}&=&\mbox{Exp}\left\{-\frac{\sqrt{\lambda}}{4}\left[1-\sqrt{1-\frac{8}{9}\frac{T_{HP}^2}{T^2}}\right]\right\}= 1-\frac{\sqrt{\lambda}}{9}\frac{T_{HP}^2}{T^2}+{\mathcal O}\left(\frac{1}{T^4}\right),\label{eq.PL-S}
 \end{eqnarray}
where $\lambda\equiv g_{\rm{YM}}^2 N=4\pi g_sN$. Similar as the free energy density (\ref{eq.f-SYM2}), it achieves only power corrections of $1/T^2$.
(\ref{eq.l-SYM1}) and (\ref{eq.PL-S}) shows clearly the sharp difference between the strong and weak coupling regime. From (\ref{eq.l-SYM1}) one can check that corrections up to ${\mathcal O}(1/T^5)$ decreases the Polyakov loop only by $12\%$ at $T_H$, and a long power series is needed to recover the exact value $1/2$. Contrary to it, with the first two terms the strong coupling expression (\ref{eq.PL-S}) is almost exact in the whole deconfined phase. In FIG.~\ref{fig:Polyakov-SYM} we plot the different behavior of the Polyakov loop at weak and strong coupling, based on the truncated solution (\ref{eq.rho1-trunc}) and the exact expression in (\ref{eq.PL-S}). In the latter we have fixed $\lambda=36 \log ^2 2$, so that the Polyakov loop approaches $1/2$ at $T_{HP}$. Notice that with such a value for $\lambda$ the first two terms in (\ref{eq.PL-S}) gives ${\mathcal L}(T_{HP})\sim 0.54$, very close to the exact value.



\section{Summary and discussion}
Following the matrix formalism in \cite{Aharony:2003sx} for large-$N$ gauge theories on a sphere, we have derived explicitly the high-temperature expansion of the Polyakov loop and the free energy at zero coupling. If one abandons the gauge-invariance constraint and approximates the eigenvalue distribution by the $\delta$-function, the previous result obtained with the Heat-Kernel method is recovered. With such a constraint kept, the eigenvalues are distributed within a small arc of the unit circle at high temperature. The open angle of the arc is determined completely by the single-particle partition functions. By dimension analysis, one finds the square of the angle vanishes as $1/T^3$ when $T\to \infty$, for any gauge theories at large $N$. In turn, all the moments $\rho_n$ of the eigenvalue density function, including the Polyakov loop $\rho_1$, achieve corrections starting from $1/T^3$. While the coefficients of such corrections for $\rho_n$ differ among different theories, they combine to give a universal term, $T/4$, in the free energy. Such a term is in some sense similar to the constant term in the expansion of the single-partition function $z_{V4}(x)$ for the vector field on the sphere.

The previous result from the Heat-Kernel approach appears to be very similar as the corresponding result at strong coupling obtained through the gauge/string duality. The similarity indicates a smooth interpolation between the weak and strong coupling regime. Since such a similarity is spoiled when the gauge-invariance constraint is imposed, the smoothness of the evolution with coupling needs to be reconsidered. To obtain the exact behavior we must extend all the calculations at zero coupling to include the finite coupling corrections. Some calculations with a small coupling has been given in~\cite{Aharony:2003sx}, with the emphasis on the phase structure. For that the perturbative corrections are crucial because at the phase transition the quadratic term of $\rho_1$ vanishes. Corrections of different signs give different phase structure at nonzero coupling. The general structure of the perturbative contributions are also shown. Roughly, higher order perturbative terms appear with more traces of the holonomy $U$ to some power. If one consider only double-trace operators in the integrand of the partition function, i.e., only perturbative corrections to $a_n$, our derivation is still valid and the expansion pattern for $\rho_n$ does not change at all. The coefficient of each term achieves perturbative corrections, in accordance with those of $a_n$. However, in the general case multi-trace terms appear, even at order $\lambda$~\cite{Aharony:2003sx,Aharony:2005bq,Aharony:2006rf}. The solutions of the matrix model with such terms are still not so clear as far as we know, but we can still make some speculations with the present results.
We have seen that the quadratic corrections to the moments $\rho_n$ are vanishing at zero coupling. Such terms may  appear when the multi-trace terms at nonzero coupling are included. However, the $1/T^3$ corrections to $\rho_n$ are finite at zero coupling. These finite terms can not be compensated by finite order perturbative corrections, as long as the coupling is small enough. Therefore such $1/T^3$ terms continue to exist in the small coupling regime. For ${\mathcal N}=4$ SYM at strong coupling, the situation is contrary. The quadratic correction to the Polyakov loop is finite and the odd terms are vanishing. In order to interpolate these two limits smoothly, the only possibility is that the theory could generate quadratic correction at small coupling, and odd terms at large but finite coupling. It would be interesting to check such a possibility directly. For example, one could consider the deformation of the global AdS black hole solution when higher derivative terms are included, in parallel with that in \cite{Gubser:1998nz}. However, we could give two arguments that this will not happen in general. First, the lattice simulation in pure gauge theory indicates that the odd power terms in the Polyakov loop do not show up at strong coupling. Secondly, the linear term $T/4$ in the free energy is universal and does not depend on the underlying dynamics. One can hardly believe that such a universal term could evolve dynamically in each theory to zero in the strong coupling limit. Based on these facts, we believe it is quite possible that a strong/weak coupling phase transition occurs for large-$N$ gauge theories on a compact manifold. Such a phase transition may even happen not just for the maximal supersymmetric theories as conjectured in \cite{Li:1998kd,Gao:1998ww}, since the present results are obtained in general. With such a phase transition, the differences between the strong and weak coupling regime could be naturally explained.

The Hagedorn transition studied here is in many sense similar to the Gregory-Laflamme transition in gravity theory~\cite{Gregory:1994bj}. The relation between them has been studied intensively, see for example~\cite{Aharony:2004ig,Mandal:2011ws}. The distribution of the Polyakov loop eigenvalues is related to the black brane distribution on a T-dual time circle on the gravity side. The Gregory-Laflamme transition is from a phase with uniform black string/brane to that of nonuniform black string/brane or localized black hole. An extension of the present discussion to that for the Gregory-Laflamme phase transition would be interesting, and possibly help understanding the present results.

\section*{Acknowledgments}
The work was initiated while F.Z. participated in the 7th Edition of the International Workshop on Quantum Chromodynamics (QCD@Work 2014), 16-19 June, Giovinazzo (Bari - Italy). F.Z. would like to thank the organizers for warm hospitality. The work is partially supported by the National Natural Science Foundation of China under Grant No. 11405065 and No. 11445001.

\end{document}